\documentclass[aip,reprint,graphicx]{revtex4-1}
\usepackage{graphicx}
\usepackage{xcolor}
\usepackage{pifont}
\usepackage{enumitem}
\usepackage{textcomp}
\setitemize{noitemsep,topsep=0pt,parsep=0pt,partopsep=0pt}
\usepackage{amsmath}

\begin{document}

\title{Soft matter crystallography---complex, diverse, and new crystal structures in condensed materials on the mesoscale}

\author{Julia Dshemuchadse}
\email{jd732@cornell.edu}
\affiliation{Department of Materials Science and Engineering, Cornell University, Ithaca, NY 14853, USA}

\date{\today}

\begin{abstract}

An increasing variety of crystal structures has been observed in soft condensed matter over the past two decades, surpassing most expectations for the diversity of arrangements accessible through classical driving forces. 
Here, we survey the structural breadth of mesoscopic crystals---formed by micellar systems, nanoparticles, colloids \textit{etc.}---that have been observed both in soft matter experiments and coarse-grained self-assembly simulations. 
We review structure types that were found to mimic crystals on the atomic scale, as well as those that do not correspond to known geometries and seem to only occur on the mesoscale. 
While the number of crystal structure types observed in soft condensed matter still lags behind what is known from hard condensed matter, we hypothesize that the high tunability and diversity of building blocks that can be created on the nano- and microscale will render a structural variety that far exceeds that of atomic compounds, which are inevitably restricted by the ``limitations'' imposed by the periodic table of elements and by the properties of the chemical bond. 
An infusion of expertise in structural analysis 
from the field of crystallography into the soft condensed matter community will establish the common language necessary to report, compare, and organize 
the rapidly accruing structural knowledge gathered from simulations and experiments. 
The prospect of new materials created in soft matter and new, length-scale-spanning insights into the formation of ordered structures in both hard and soft condensed matter 
promise exciting new developments in the area of self-assembled mesoscale materials.
\end{abstract}

\maketitle


\section*{Introduction}

\begin{figure}
\centering
\includegraphics[width=\columnwidth]{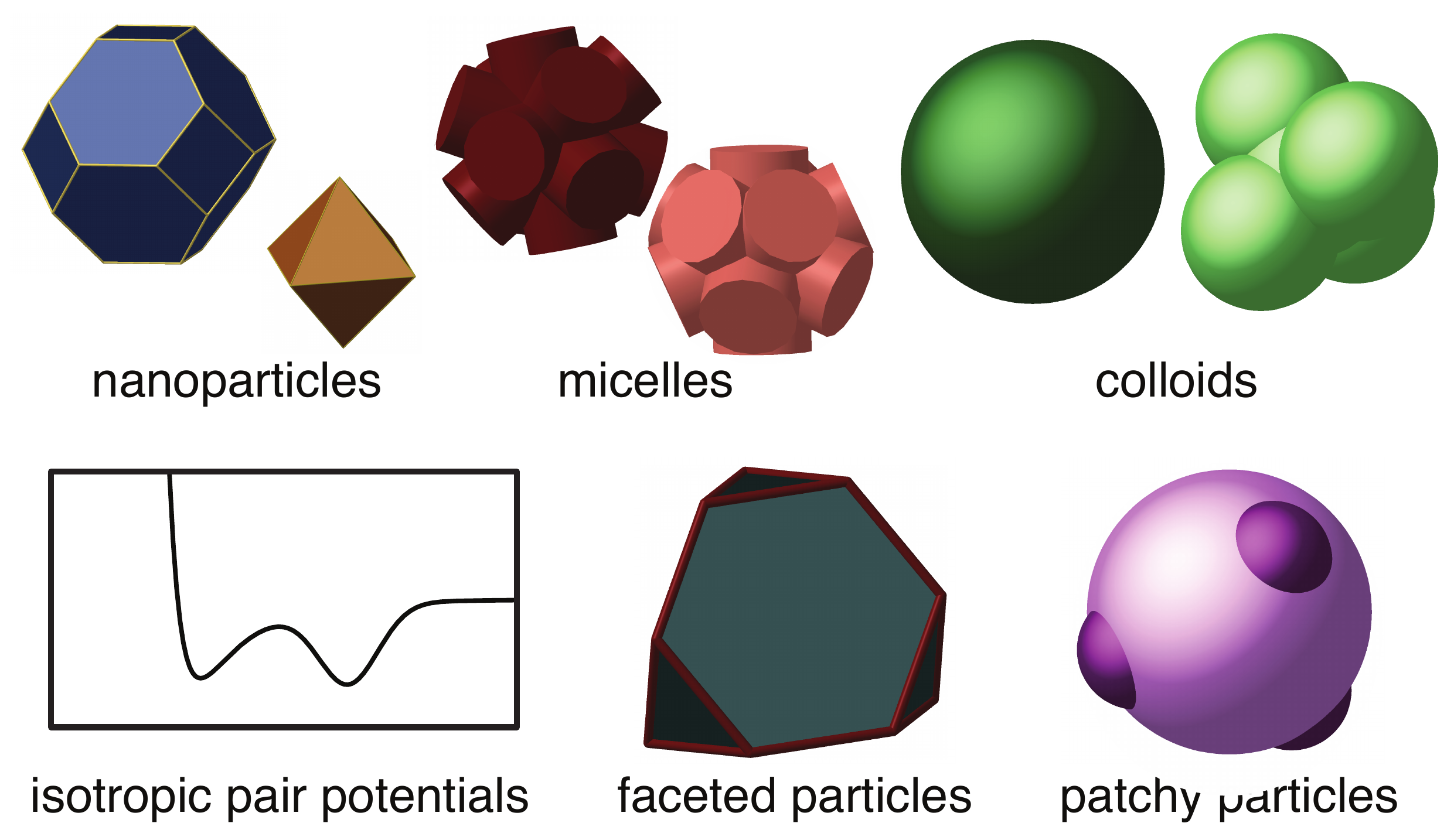}
\caption{\textit{Top:} Types of mesoscale particles: nanoparticles or nanodots, micelles, colloids. 
\textit{Bottom:} Types of computational models most commonly used to simulate mesoscale systems: isotropic pair potentials, faceted (often: hard) polyhedra, anisotropic patchy particles.}
\label{compmodels}
\end{figure}

It has been argued that complex building blocks beget complex assemblies---that the creation of complex materials can only be achieved through intricate and highly specific interactions.\cite{Lukatsky2006} 
Increasingly, however, complex structures have been shown to arise on the mesoscale from deceptively simple particle systems.\cite{Boles2016} 
In experimental studies, a growing number of various crystal structures can be formed from 
nanoparticles, micellar systems, and colloids, 
whereas most computational studies on the mesoscale use isotropic pair potential, anisotropic faceted polyhedral shapes, or patchy particle models (see Fig.~\ref{compmodels}).

Many of the structures formed in experiments and simulations of particulate matter on the mesoscale are 
being found to form in particulate matter on the mesoscale 
and are geometrically equivalent to crystals on the atomic length scale. 
While highly complex crystal structures on the atomic scale---\textit{e.g.}, quasicrystals and complex intermetallics\cite{Steurer2009,Dshemuchadse2012}---are stabilized by electronic interaction and forces at the quantum level, 
mesoscale crystal structures are formed due to classical interactions. 
The observation of various intricate crystal structures self-assembling from 
simple numerical simulations\cite{HajiAkbari2009,Engel2015}---operating only through classical thermodynamics---is 
therefore a logical extension of the experimental detection of such structures in soft condensed matter. 

Another factor sets soft matter apart from hard matter: 
the highly discrete choice of building blocks on the atomic scale---dictated by the periodic table---is in stark contrast to the dizzying number of parameters at the disposal of soft materials 
scientists and engineers.\cite{Glotzer2007,vanAnders2014b}
The enormous tunability of building blocks on the mesoscale implies that there is 
no reason to restrict ourselves to targeting structures known from compounds on the atomic scale when creating new ordered mesoscale materials.

In order to access the full diversity of feasible crystal structure types as design targets in soft condensed matter, we must know which structures are technically possible. 
Currently, all 
those that we know of and can easily search for through existing databases---such as the \textit{Inorganic Crystal Structure Database} (ICSD)\cite{Allmann2007} and \textit{Pearson's Crystal Data}\cite{Villars2020}---are atomic or molecular structures, whose geometries are dictated by the properties of the chemical bond. 
If we aim to design 
completely novel structures, we must invent 
new ways of looking for them, 
while adhering to the tried and tested methods from the crystallographic playbook to describe and analyze said structures.


\section*{Background: structural complexity and variety}

Within the traditional realm of crystallography, where all ordered structures can be represented with 230 space groups and captured in a parallelepiped that serves as an immutable repeat unit, the size of that unit cell can reasonably be used to measure the complexity of the crystal structure that it describes.\cite{Dshemuchadse2012} 
Here, this is expressed in 
a notation for different crystal structure types 
that consists 
of the Pearson symbol, which specifies a structure's Bravais lattice and the number of atoms per unit cell (\textit{e.g.}, $cF4$ for the face-centered cubic close-packing of spheres, which contains 4 atoms per unit cell), as well as a prototypical compound that exhibits the respective crystal structure type (\textit{e.g.}, $cF4$-Cu in the case of the cubic-close sphere packing; more details on the crystallographic notation used here are indicated in the Appendix). 
(Additional variations of 
structural complexity measures
include using only the primitive unit cell's content, as opposed to the often centered unit cells used as standard settings, or using the contents of only the asymmetric unit, \textit{i.e.}, the number of Wyckoff positions, which are then multiplied by rotations, rotoinversions, glide planes, and screw axes to populate the unit cell. 
Alternatively, the complexity of crystal structures can be expressed \textit{via} a topological approach based on information theory.\cite{Krivovichev2012a,Krivovichev2014})

While all periodic crystals can be characterized with space groups and their unit cell size therefore measured accordingly, not all crystals are periodic.\cite{IUCr1992} 
Aperiodic crystals---such as quasicrystals or incommensurately modulated crystal structures---can only be described by periodic units that possess additional dimensions compared with the spatial dimensions of the crystal itself,\cite{Janssen2007,vanSmaalen2007,Steurer2009} 
making it impossible to compare their unit cell contents to those of classic, three-dimensional ones. 

An additional degree of complexity in crystalline structures is that of intrinsic disorder.\cite{Welberry2010,Keen2015} While ``crystals'' are inherently mostly ordered structures, disorder can still prevail within them either in statistical or correlated manners. 
In some cases this is the result of incomplete crystallization, but correlated or partial disorder can also be an inherent feature of a crystal structure and---while difficult to decipher---it can even present a tunable design parameter.\cite{Simonov2020}  

Some of these dimensions of structural complexity are difficult to enumerate: 
aperiodic crystal structures and those with correlated disorder defy the standard means of reporting crystal structures and are therefore not easily encompassed and represented in databases. 
Despite these gaps regarding 
extreme structural complexity, however, the sheer variety of crystal structures known from compounds on the atomic scale is dizzying: 
the ICSD, for example, contains over 200,000 entries, a majority of which can be assigned to over 9,000 crystallographically distinct structure types.\cite{Zagorac2019} 
It is important to keep in mind that these structure types belong to different materials classes, some of which are likely better design targets for mesoscale systems than others, \textit{e.g.}, the over 2,000 structure types in which intermetallic compounds have been found to occur.\cite{Dshemuchadse2015a} 


\section*{Review: complex crystal structures in hard condensed matter}

Highly complex structures on the atomic scale have been known for over half a century: 
Samson described a variety of intermetallic compounds that exhibit unit cells containing over a thousand atoms,\cite{Samson1962,Samson1965,Samson1967} 
and the largest known unit cell in intermetallic systems was found to contain more than 20,000 atoms.\cite{Weber2009,Conrad2009} 
While uncommon, complex intermetallics have been observed in a large variety of systems\cite{Dshemuchadse2011}---containing different chemical elements, featuring different types of chemical bonding, and hugely varying stoichiometries---and they make up around 2\%\ of intermetallic compounds generally\cite{Dshemuchadse2012}.
Binary and ternary intermetallics also constitute the systems where quasicrystalline structures were first discovered\cite{Shechtman1984} and later found to be all but ubiquitous.\cite{Steurer2009,Steurer2018} 
Another system discovered 
to exhibit quasicrystallinity on the atomic scale is a two-dimensional oxide in the Ba--Ti--O system (deposited on Pt$(111)$ substrates).\cite{Forster2013,Zollner2019} 

Given the structural simplicity of most pure metals, which tend to adopt sphere packing-type structures, 
intermetallic systems might be the most unsuspecting and therefore most surprising candidates for producing highly complicated crystal structures. 
However, compounds of extreme complexity are well-known in other systems, too; 
oxides in particular provide a rich landscape of structural variety. 
One class of structures, for example, consists of intergrowths of multiple structural elements at a variety of compositions; such related but distinct structure types can be understood as a homologous series and comprise theoretically infinitely many individual compounds, \textit{e.g.}, Ruddlesden-Popper phases.\cite{Sharma1998} 
Another category of oxides with complex structures are framework compounds: zeolites, for example, are minerals that occur in a huge variety of crystal structures\cite{McCusker2007,Baerlocher2007,IZAdatabase}---with even more hypothetical frameworks\cite{Earl2006}---and 
new structures are continually being discovered.\cite{Li2015} 
Other compounds with complex structures are, for example, clathrates, in which large cations---atomic or molecular in nature---occupy the even larger cages in a host structure. The host structures are typically formed by covalently-bonded elements such as silicon---in intermetallic clathrates\cite{Baitinger2014}---or by water molecules in clathrate hydrates.\cite{Pauling1952} 
In the realm between 
organic and inorganic crystal structures lie more categories of framework compounds: metal-organic frameworks (MOFs)\cite{OKeeffe2012} 
and more recently discovered covalent-organic frameworks,\cite{Cote2005} 
which also represent a dizzying variety of structure types and tunable geometrical and chemical properties. 
In a nutshell: complexity is present in crystal structures on the atomic scale in a multitude of ways 
independent of the building blocks' complexity. 


\section*{Review: complex crystal structures in soft condensed matter}

\begin{figure*}
\centering
\includegraphics[width=\textwidth]{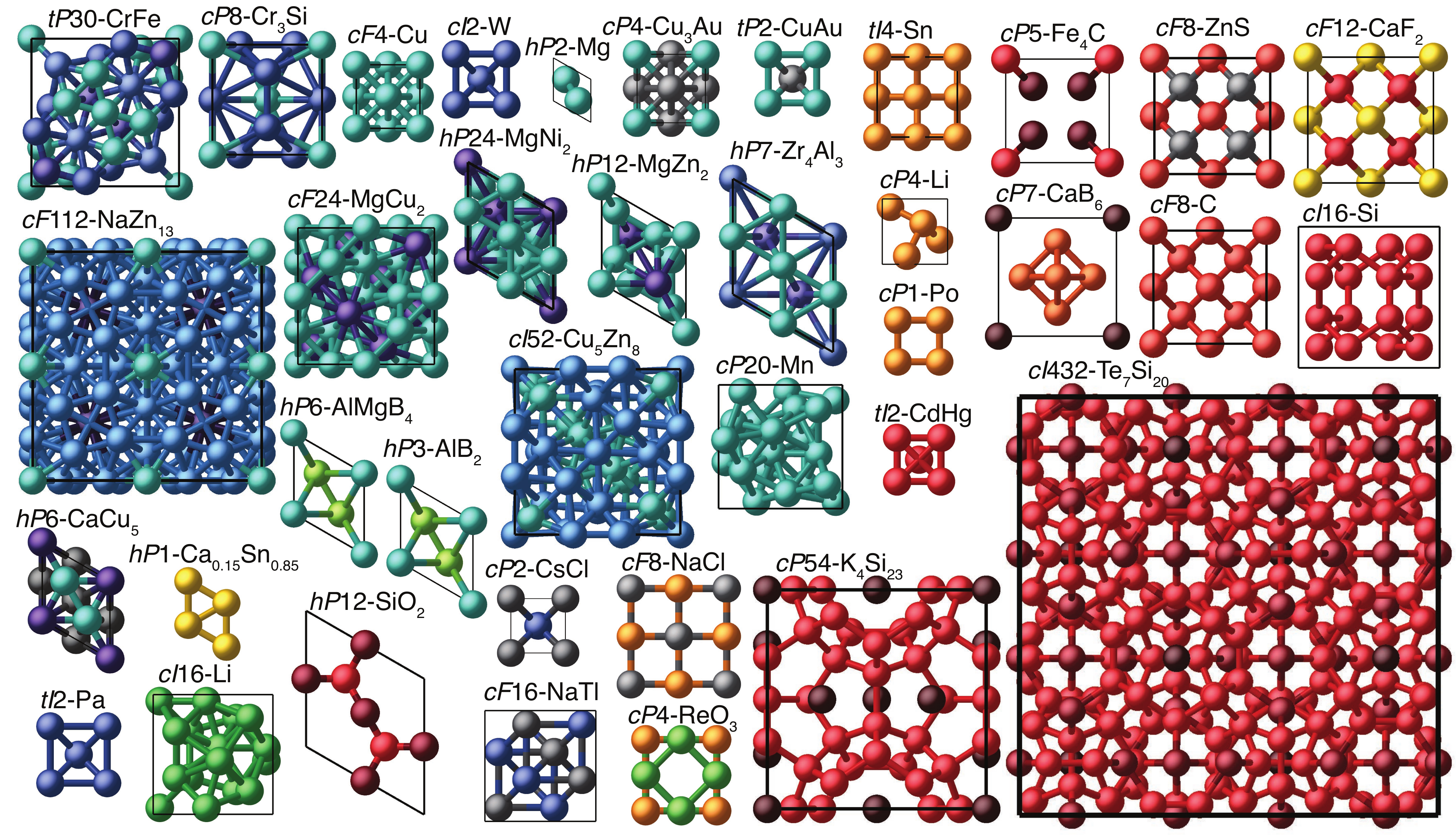}
\caption{Known crystal structure types that were found in experimental and computational studies on the mesoscale. 
Particle positions are colored according to their coordination numbers (with gray particles indicating Wyckoff positions with identical coordination numbers occupied with distinct particle types): 
maroon = 0--2, 
red = 4, 
orange = 5--6, 
yellow = 8, 
green = 9--11,
cyan = 12, 
blue = 13--15,
purple = 16--24. 
Images are generated using CrystalMaker\textsuperscript{\textregistered}.\cite{CrystalMaker}}
\label{unitcells}
\end{figure*}

Similarly to metals, soft matter systems were initially thought to mainly form simple structures. 
While the self-assembly of colloidal spheres into sphere packings was observed early and widely\cite{Wood1957,Alder1957,Pusey1989,Zhu1997}---and 
the same structure types have been observed in a growing variety of systems, such as 
functionalized nanoparticles ($cF4$-Cu, $cI2$-W, $hP2$-Mg)\cite{Macfarlane2010,Macfarlane2011}---an 
increasing diversity of structures has been uncovered 
over the past 15 years (see Tab.~\ref{structuretypes} and Fig.~\ref{unitcells}) 
through varying particle attributes such as explicit interaction potential and anisotropic particle shape. 
In addition to these attributes, however, 
there are other parameters that open up a high-dimensional design space in mesoscale materials,\cite{Glotzer2007,vanAnders2014b} 
and have lead to the discovery of ever more 
structures. 
Superstructures of sphere packings, for example, have been observed in binary systems of nanoparticles, such as $tP2$-CuAu and $cP4$-Cu$_3$Au---both derivatives of the cubic-close sphere packing $cF4$-Cu.\cite{Shevchenko2006} 

\begin{table*}
\scriptsize
\setlength{\tabcolsep}{2pt}
\caption{Known crystal structure types that were found in experimental studies on the mesoscale, and in related coarse-grained simulation studies.
The experimental mesoscale systems include: colloids (\textit{e.g.}, spherical shape (spheres), binary systems (2)), nanoparticles (NP; \textit{e.g.}, one-component (1), binary (2), ternary (3); multivalent (MV) systems; binary DNA-functionalized nanoparticles (2-DNA)), micelle-forming systems (\textit{e.g.}, block copolymers (BCP), dendrimers (dendr.), surfactants (surf.)); as well as shape amphiphiles (SA). 
The model systems used in simulations include 
isotropic particles (iso.; \textit{i.e.}, hard spheres (HS) or polydisperse sphere systems (PS), functionalized hard spheres (fHS), or point particles interacting \textit{via} Lennard Jones (LJ) or other isotropic pair potentials (IPP)) 
and anisotropic particles (aniso.; \textit{e.g.}, anisotropic polyhedra (AP; as well as corresponding binary systems: 2-AP), sphere clusters (SC), combinations of spherical clusters and spheres (SC+S), or patchy particles (PP)).}
\label{structuretypes}
\begin{tabular}{r@{-}llc|ccc|cc}
\hline\hline
\multicolumn{4}{c|}{\textbf{Structure type}} & \multicolumn{3}{c|}{\textbf{Experimental studies}} & \multicolumn{2}{c}{\textbf{Simulation studies}} \\
\multicolumn{2}{l}{Pearson symbol} & space group & colloquial name & colloids & nanoparticles & micelles & iso. & aniso.\\
\hline\hline
\multicolumn{9}{l}{\textit{Sphere packings}}\\
\hline
$cF4$&Cu & $Fm\bar{3}m$ & \textit{ccp} / ``\textit{fcc}'' / A1 & spheres\cite{Pusey1989,Zhu1997} & 2-DNA\cite{Macfarlane2010,Macfarlane2011} &  & LJ, HS\cite{Wood1957,Alder1957} & AP\cite{Damasceno2012a}; SC\cite{Marson2019} \\ 
$cI2$&W & $Im\bar{3}m$ & ``\textit{bcc}'' / A2 & spheres & 1\cite{Henzie2012}; 2-DNA\cite{Macfarlane2011} & BCP\cite{Lee2010,Lee2014,Chanpuriya2016}; surf.\cite{Yue2016,Kim2017b} & LJ & AP; SC\cite{Marson2019} \\
$hP2$&Mg & $P6_3/mmc$ & \textit{hcp} / A3 & spheres & 2-DNA\cite{Macfarlane2011} & BCP\cite{Chanpuriya2016} & LJ & AP \\
$tP2$&CuAu & $P4/mmm$ & L1$_0$ & & 2\cite{Shevchenko2006} & &  &  \\ 
$cP4$&Cu$_3$Au & $Pm\bar{3}m$ & L1$_2$ & & 2\cite{Shevchenko2006} & &  &  \\ 
$tI2$&Pa & $I4/mmm$ & A$_\text{a}$ &  &  &  & IPP\cite{Dshemuchadse2021} & \\
\hline
\multicolumn{9}{l}{\textit{Frank-Kasper phases}}\\
\hline
$cF24$&MgCu$_2$ & $Fd\bar{3}m$ & C15 (Laves) & 2\cite{Ducrot2017} &  & surf.\cite{BaezCotto2018} & PS\cite{Lindquist2018,Dasgupta2020} & \\ 
$hP12$&MgZn$_2$ & $P6_3/mmc$ & C14 (Laves) &  & 1\cite{Hajiw2015}; 2\cite{Shevchenko2006} & surf.\cite{BaezCotto2018} &  PS\cite{Lindquist2018} & \\ 
$hP24$&MgNi$_2$ & $P6_3/mmc$ & C36 (Laves) &  & 2\cite{Shevchenko2006} & &  & \\
$cP8$&Cr$_3$Si & $Pm\bar{3}n$ & A15 & & 2-DNA\cite{Macfarlane2011} & BCP\cite{Cho2004,Chanpuriya2016}; surf.\cite{Huang2015,Yue2016,Kim2017b}; dendr.\cite{Balagurusamy1997,Ungar2005} & IPP\cite{Engel2015,Lindquist2018b} & AP\cite{Damasceno2012a} \\
$tP30$&CrFe & $P4_2/mnm$ & $\sigma$-phase / D8$_\text{b}$ & &  & BCP\cite{Lee2010,Lee2014,Schulze2017}; surf.\cite{Yue2016,Kim2017b} & IPP\cite{Dshemuchadse2021} & AP\cite{Damasceno2012a} \\
$hP7$&Zr$_4$Al$_3$ & $P6/mmm$ & Z-phase &  & & SA\cite{Su2019} & IPP\cite{Dshemuchadse2021} & SC\cite{Marson2019} \\ 
\multicolumn{4}{l|}{dodecagonal quasicrystals (12-QC)} & $+$\cite{Fischer2011} & 2\cite{Talapin2009,Ye2017} & BCP\cite{Zhang2012,Chanpuriya2016}; surf.\cite{Yue2016}; dendr.\cite{Zeng2004} & IPP\cite{Dotera2014} &  \\ 
\hline
\multicolumn{9}{l}{\textit{More intermetallic structure types}}\\
\hline
$cI52$&Cu$_5$Zn$_8$ & $I\bar{4}3m$ & $\gamma$-brass / D8$_2$ &  & &  & IPP\cite{Zetterling2000} & AP\cite{Damasceno2012a}; SC\cite{Marson2019} \\ 
$cP20$&Mn & $P4_132$ & $\beta$-Mn / A13 &  &  & & IPP\cite{Elenius2009} & AP\cite{Damasceno2012a}; SC\cite{Marson2019} \\ 
$cF112$&NaZn$_{13}$ & $Fm\bar{3}c$ & D2$_3$ &  & 2\cite{Shevchenko2006} &  & PS\cite{Bommineni2019} & \\
$hP6$&CaCu$_5$ & $P6/mmm$ & D2$_\text{d}$ &  & 2\cite{Shevchenko2006} & & & \\
$hP3$&AlB$_2$ & $P6/mmm$ & C32 & +DNA\cite{Wang2015} & 2\cite{Shevchenko2006}; 2-DNA\cite{Macfarlane2011} &  &  & \\
$hP6$&AlMgB$_4$ & $P6/mmm$ & &  & 3\cite{Evers2009} &  & & \\
$cP7$&CaB$_6$ & $Pm\bar{3}m$ & D2$_1$ &  & 2\cite{Shevchenko2006,Ye2011} &  &  & \\
\hline
\multicolumn{9}{l}{\textit{Significant ionic contributions (salts, Zintl phases, \textit{etc.})}}\\
\hline
$cF8$&NaCl & $Fm\bar{3}m$ & rocksalt / B1 &  & 2\cite{Shevchenko2006}; MV\cite{Macfarlane2011} &  & fHS\cite{Tkachenko2002} & \\
$cP2$&CsCl & $Pm\bar{3}m$ & B2 & $+$\cite{Wang2015} & 2-DNA\cite{Macfarlane2011} &  & fHS\cite{Tkachenko2002} & \\
$cF16$&NaTl & $Fd\bar{3}m$ & B32 &  & MV\cite{Cigler2010,Wang2017} & surf.\cite{BaezCotto2018} &  & \\
$cF8$&ZnS & $F\bar{4}3m$ & zincblende / B4 & & & & fHS\cite{Tkachenko2002} & \\
$cF12$&CaF$_2$ & $Fm\bar{3}m$ & fluorite / C1 & & & & & 2-AP\cite{Cadotte2016} \\
$cP4$&ReO$_3$ & $P23$ & D0$_9$ & & & & & SC+S\cite{Zanjani2016} \\
$hP12$&SiO$_2$ & $P6_3/mmc$ & $\beta$-tridymite / C10 & & & & & SC+S\cite{Zanjani2016} \\
\hline
\multicolumn{9}{l}{\textit{Low coordination numbers}}\\
\hline
$cP1$&Po & $Pm\bar{3}m$ & A$_\text{h}$ & cubes\cite{Rossi2011} & 1\cite{Henzie2012}; MV\cite{Macfarlane2011} &  & IPP\cite{Rechtsman2006,Jain2013} & AP\cite{Damasceno2012a,vanDamme2017} \\ 
$tI4$&Sn & $I4_1/amd$ & $\beta$-tin / A5 &  &  &  & IPP\cite{Dshemuchadse2021} & AP\cite{Damasceno2012a} \\ 
$cI16$&Li & $I\bar{4}3d$ & &  & 1\cite{Henzie2012} &  &  & AP\cite{Damasceno2012a,Damasceno2012b} \\
$cI16$&Si & $Ia\bar{3}$ & &  &  &  & IPP\cite{Engel2015,Dshemuchadse2021} & \\
$tI2$&CdHg & $I4/mmm$ & &  &  &  & IPP\cite{Dshemuchadse2021} & \\
$hP1$&Ca$_{0.15}$Sn$_{0.85}$ & $P6/mmm$ & &  & &  & IPP\cite{Dshemuchadse2021} & \\
$cP4$&Li & $P4_132$ & &  &  & & IPP\cite{Engel2015,Dshemuchadse2021} & \\
$cF8$&C & $Fd\bar{3}m$ & diamond / A4 & tetrapods\cite{He2020} &  & & IPP\cite{Rechtsman2007,Jain2013,Bertolazzo2016} & AP\cite{Damasceno2012b,Damasceno2012a}, PP\cite{Zhang2005,Romano2009,Noya2010,Patra2018} \\
$cP54$&K$_4$Si$_{23}$ & $Pm\bar{3}n$ & clathrate-I &  &  & & IPP\cite{Engel2015,Spellings2018,Dshemuchadse2021} & PP\cite{Noya2019} \\
$cI432$&Te$_{7+x}$Si$_{20-x}$ & $Fd\bar{3}c$ &  &  &  & & & AP\cite{Lee2019} \\
$cP5$&Fe$_4$C & $P\bar{4}3m$ & &  & 2\cite{Shevchenko2006} & & & \\
\hline\hline
\end{tabular}
\end{table*}
\normalsize

The next most intuitive type of structures that we might expect to observe in assemblies of isotropic mesoscale particles are binary sphere packings, known from the structures of metallic compounds and characterized by their 12-, 14-, 15-, and 16-neighbor local environments---the Frank-Kasper phases.\cite{Frank1958,Frank1959,Travesset2017b,Reddy2019} 
These structure types are also called ``topologically close packed'', and they can be considered to be packings of (distorted) tetrahedral interatomic voids. 
The most common Frank-Kasper phases known on the atomic scale are also ubiquitous on the mesoscale:
$cP8$-Cr$_3$Si (or ``A15 phase'') was found to form in systems of dendrimer micelles,\cite{Balagurusamy1997,Ungar2005} 
block copolymers,\cite{Cho2004,Chanpuriya2016} 
monomers with rigid tetrahedral frames at the core and polyhedral molecular units at the vertices,\cite{Huang2015} 
surfactants,\cite{Yue2016,Kim2017b} 
and binary systems of DNA-functionalized nanoparticles.\cite{Macfarlane2011} 
In simulations, it has been found in systems of hard faceted polyhedra\cite{Damasceno2012a} 
as well as isotropic multi-well pair potentials.\cite{Engel2015}
$tP30$-CrFe (or ``$\sigma$-phase'', which is a dodecagonal quasicrystal approximant) was also found in systems of block copolymer micelles\cite{Lee2010,Lee2014,Schulze2017} 
and surfactants,\cite{Yue2016,Kim2017b} 
as well as in simulations of hard polyhedra\cite{Damasceno2012a} 
and isotropic pair potentials.\cite{Engel2008b,Dshemuchadse2021} 
(Dodecagonal quasicrystals themselves have been found in systems of 
dendrimers,\cite{Zeng2004} 
star polymers,\cite{Hayashida2007} 
block copolymers,\cite{Zhang2012,Chanpuriya2016} 
surfactants,\cite{Yue2016} 
colloids,\cite{Fischer2011} 
mesoporous silica,\cite{Xiao2012} 
binary systems of nanoparticles,\cite{Talapin2009,Ye2017} 
as well as Monte Carlo simulations of repulsive particles interacting \textit{via} multi-step interaction potentials.\cite{Dotera2014}) 
The less-common $hP7$-Zr$_4$Al$_3$ structure has been 
found in systems of shape amphiphiles\cite{Su2019}, 
in simulations of softly repulsive clusters of dodecahedral spheres,\cite{Marson2019} 
and in particles interacting \textit{via} isotropic pair potentials.\cite{Dshemuchadse2021} 
A very common subtype of Frank-Kasper phases are Laves phases, which can be understood as layered structures composed of only 12- and 16-coordinated local environments with $AB_2$-type compositions.\cite{Stein2021} 
Two Laves phases are particularly common: the cubic Laves phase $cF24$-MgCu$_2$ (or ``C15'' phase), which has been observed 
in mixtures of spherical colloids and tetrahedrally-shaped colloidal clusters,\cite{Ducrot2017} and the hexagonal Laves phase $hP12$-MgZn$_2$ (or ``C14'' phase), 
which was found in Au nanoparticle superlattices\cite{Hajiw2015} and in binary nanoparticle systems;\cite{Shevchenko2006} 
both of these structures have also been observed 
in surfactant micelles\cite{BaezCotto2018} 
and in simulations of polydisperse particle systems\cite{Lindquist2018} and slightly soft repulsive binary particle systems;\cite{Dasgupta2020} 
while a third Laves phase $hP24$-MgNi$_2$ was only found in binary nanoparticle systems.\cite{Shevchenko2006} 

Other structure types that are ubiquitous in intermetallic compounds are closely related to topologically close-packed phases: 
$\gamma$-brass $cI52$-Cu$_5$Zn$_8$ has been observed in simulations of hard truncated dodecahedra\cite{Damasceno2012a} and repulsive clusters of dodecahedral spheres,\cite{Marson2019} 
in other hard shapes,\cite{Klotsa2018} 
as well as isotropic pair potentials,\cite{Zetterling2000} 
and $\beta$-manganese $cP20$-Mn has been observed in simulations of pentagonal dodecahedra, as well as some of its augmented versions,\cite{Damasceno2012a} 
and corresponding repulsive sphere clusters,\cite{Marson2019} 
and also isotropic pair potentials.\cite{Elenius2009} 
A sphere packing with a more disparate size ratio---and therefore 12- and 13-, as well as a 24-coordinated site---is $cF112$-NaZn$_{13}$, which has been observed both in binary systems of nanoparticles\cite{Shevchenko2006} and in simulations of polydisperse systems.\cite{Bommineni2019} 
In similar binary systems of nanoparticles, the $hP6$-CaCu$_5$ structure type has been observed, exhibiting 12- and 18-coordinated sites.\cite{Shevchenko2006} 

Larger discrepancies in particle size in multi-component systems can stabilize structures such as $hP3$-AlB$_2$, which has been observed in binary systems of nanoparticles,\cite{Shevchenko2006} 
as well as DNA-functionalized colloids,\cite{Wang2015} 
and $hP6$-AlMgB$_4$---a ternary superstructure of $hP3$-AlB$_2$---observed in a ternary system of nanocrystals,\cite{Evers2009} 
and $cP7$-CaB$_6$, also observed in binary nanoparticle systems.\cite{Ye2011} 
Another binary structure type stabilized in this manner is the salt-type $cP2$-CsCl, which was also observed in DNA-functionalized nanoparticles and colloids,\cite{Macfarlane2011,Wang2015} 
and in binary systems of hard spheres with attractive interactions between unlike particles.\cite{Tkachenko2002} 

The above-mentioned systems modify shape parameters of the constituent particles, in particular \textit{via} anisotropy in systems of faceted polyhedra or compound shapes such as rigid bodies of multiple spheres, or also \textit{via} the size ratio in multi-component systems. 
An even larger variety of crystal structures becomes accessible if specific interaction patterns can be implemented, which can drive the assembly of more multi-component phases \textit{via} multivalent functionalizations.
Examples of the new types of structures that can be attained in this manner are:
another salt-type structure formed by nanoparticles---$cF8$-NaCl\cite{Macfarlane2011} (also observed in binary systems of nanoparticles\cite{Shevchenko2006})---and a Zintl-phase---$cF16$-NaTl\cite{Cigler2010,Wang2017} (or B32, also called ``double-diamond''; also observed in surfactant micelles\cite{BaezCotto2018}). 
Both with nanocrystals with polymer-mediated interactions\cite{Henzie2012} 
and with DNA-functionalized nanoparticles with multivalent interactions but identical cores, 
the simple cubic structure $cP1$-Po could also be assembled;\cite{Macfarlane2011} 
the same structure type was also observed 
in binary systems of hard spheres with attractive interactions between unlike particles,\cite{Tkachenko2002} 
in systems of cubic colloids,\cite{Rossi2011,vanDamme2017} 
and in simulations of hard faceted cubes,\cite{Damasceno2012a} 
as well as isotropic pair potentials.\cite{Rechtsman2006,Jain2013} 
Binary systems of hard tetrahedra and octahedra
were observed to self-assemble into the fluorite structure type, $cF12$-CaF$_2$, typically associated with ionic bonding interactions,\cite{Cadotte2016} 
while the assembly of cubic colloidal clusters combined with spherical colloids was found to result in a structure corresponding to the $cP4$-ReO$_3$ structure type
and the assembly of tetrahedral colloidal clusters combined with spherical colloids 
formed the $\beta$-tridymite structure type $hP12$-SiO$_2$.\cite{Zanjani2016} 

While metallic structure types have high coordination numbers, \textit{i.e.}, numbers of nearest neighbors of 12 and more, the 
structures from the previous paragraphs are characterized by much lower numbers of nearest neighbors (down to 6). 
The observation of low-coordinated structures has mostly occurred in simulations, 
with few exceptions such as 
a high-pressure lithium structure $cI16$-Li, which was found both in systems of polyhedral nanocrystals with polymer-mediated interactions\cite{Henzie2012} 
and in simulations of hard polyhedra.\cite{Damasceno2012a,Damasceno2012b} 
In systems of hard polyhedra, the self-assembly of $\beta$-tin $tI4$-Sn was also observed,\cite{Damasceno2012a,Damasceno2012b} and particles that interact \textit{via} isotropic pair potentials self-assembled $tI4$-Sn, $cI16$-Si, $tI2$-CdHg, $tI2$-Pa, $hP1$-Ca$_{0.15}$Sn$_{0.85}$, and $cP4$-Li.\cite{Engel2015,Dshemuchadse2021} 

Effectively, the lowest coordination number in three-dimensionally connected crystal structures is 4, and the most highly symmetric structure type to exhibit it is that of diamond ($cF8$-C). 
Only recently have tetrahedrally-shaped colloidal clusters been found to assemble this structure on the mesoscale,\cite{He2020} 
whereas simulations have discovered diamond-type crystals in systems of patchy particles,\cite{Zhang2005,Romano2009,Noya2010,Patra2018} 
as well as in systems of 
hard truncated tetrahedra,\cite{Damasceno2012b} 
and isotropic pair potentials.\cite{Rechtsman2007,Jain2013} 
The binary variant of the diamond structure type, $cF8$-ZnS, was found in binary systems of hard spheres with attractive interactions between unlike particles.\cite{Tkachenko2002} 

Another family of crystal structure types with coordination number 4 are clathrates (which on the atomic scale consist of, \textit{e.g.}, a $sp^3$-bonded Si-network): 
the simplest type, $cP54$-K$_4$Si$_{23}$ (``clathrate I''), has been observed in isotropic pair potentials, in part in competition with other clathrate structure types.\cite{Engel2015,Spellings2018,Dshemuchadse2021} 
Similar behavior has been observed in tetrahedrally patchy particles,\cite{Noya2019} 
whereas in hard polyhedral particles the highly complex clathrate-type crystal structure $cI432$-Te$_{7+x}$Si$_{20-x}$ has been observed.\cite{Lee2019} 
In binary systems of nanoparticles, the $cP5$-Fe$_4$C structure type was observed, consisting of tetrahedral Fe$_4$C colloidal ``molecules''.\cite{Shevchenko2006} 

It should be noted that lower-coordinated structures are strictly-speaking possible, 
such as 3-connected networks (\textit{e.g.}, the gyroid or double-gyroid structures) 
2-connected columnar structures, 
or 1- or 2-connected quasi-molecular structure types. 
In polymeric systems, the gyroid structure is observed as one of infinitely extended domains with minimal surface area rather than one of 3-connected particles, which is why we chose not to include it in this overview. 
Additionally, it seems to be quite rare to see such low-coordinated structures that correspond to known structures---in particular in experimental studies of soft condensed materials---although a few without atomic equivalents are mentioned below to have occurred in simulation studies.\cite{Bertolazzo2016}

Different degrees and types of order form in soft-condensed-matter systems: 
in addition to traditional, periodic crystal structures (see Tab.~\ref{structuretypes}), 
\textit{e.g.}, 
aperiodic quasicrystals,\cite{Engel2015,Damasceno2017} 
degenerate crystals,\cite{HajiAkbari2011} 
mesophases such as liquid and plastic crystals\cite{Damasceno2012a}
are observed. 


\section*{Origins of structural complexity on the mesoscale}

Many of the complex structures that arise on the mesoscale are due to the interplay of multiple, competing interactions that complicate crystal structure formation, 
such as local \textit{vs.\ }global packing in anisotropic particles\cite{Cersonsky2018a}
and soft corona \textit{vs.\ }hard core interactions in isotropic particles.\cite{Moffitt2013} 

Between particles with anisotropic shapes, implicit directional entropic forces emerge that govern the formation of ordered structures.\cite{vanAnders2014a,vanAnders2014b,Harper2019} 
When different local motifs compete with one another, crystallization can be suppressed,\cite{Teich2019} 
and similarly to hard condensed matter, phase transitions between different crystal structures can be induced \textit{via} thermodynamic 
parameters such as pressure or \textit{via} particle properties such as shape.\cite{Cersonsky2018,Du2017,Du2020} 
Transitions between the disordered and ordered states during the self-assembly process of particles with anisotropic shapes 
are often preceded by the emergence of distinct local motifs\cite{Thapar2014} 
and can feature different behaviors, such as 
nucleation and growth,\cite{Sharma2018b,Newman2019} 
the formation of disordered precursor phases,\cite{Lee2019} 
the formation of precursor mesophases,\cite{John2004,Agarwal2011,Karas2019} 
or
the gradual emergence of ordered domains.\cite{Sharma2018a} 

Most mesoscopic interactions are largely isotropic, 
be it due to the inherent sphericity of the particles themselves (micelles or non-patchy colloids) or due to the ``shielding'' of anisotropic features such as shape due to a corona with which, for example, nanoparticles are often functionalized in order to enable their self-assembly into ordered structures (\textit{e.g.}, with DNA or polymers). 
For diblock copolymers, 
the formation of Frank-Kasper structures---with unequal domain volumes---instead of simple sphere packings seems to result 
from the relative stiffness of the corona, 
leading to the formation of aspherical domains.\cite{Reddy2018} 
Early work discussed how 
star polymers and polymer coils can be described as soft colloidal particles.\cite{Likos1998,Louis2000,Bolhuis2001,Likos2006}
Inorganic nanoparticles that are functionalized with polymer brushes, for example, are said to interact \textit{via} effective potentials that exhibit features of competing interactions---of hard, repulsive core interactions and soft, attractive interactions of the particle coronas.\cite{Moffitt2013,Bretonnet2019,Fernandes2013} 
While some studies have explicitly described experimental systems of soft matter particles with isotropic pair potentials,\cite{Likos2002,Stieger2004} 
effective potentials between mesoscopic particles\cite{Likos2001,Malescio2007} 
are often simply assumed to correspond to certain isotropic pair potentials, 
and establishing a direct connection between specific particle systems and pair interactions that mimic their physical behavior remains a challenge.\cite{Jain2013,Jain2014a} 
The radial complexity of isotropic pair potentials can be viewed as an approximation of 
the complexity of potentials of mean force in many-body systems,\cite{Rechtsman2007} 
and isotropic pair potentials have been found to 
achieve agreement between coarse-grained self-assembly simulations and soft-matter experiments.\cite{Watzlawek1999,Gottwald2004}


\section*{Review: discovery of new crystal structures on the mesoscale}

While the increasing structural diversity that has been realized both experimentally and in simulations is exciting, 
the discovery of completely new crystal structures on the mesoscale 
that have no atomic equivalents 
opens up an even larger realm of structural possibility.
Our recent 
study of a large variety of isotropic pair interactions revealed the self-assembly of particles into previously unknown structure types across a wide range of complexities (from $hP1$ to $cI100$), symmetries (cubic, tetragonal, hexagonal, and orthorhombic), 
and effective ``chemical bonding states'', expressed \textit{via} their coordination numbers ranging from \textit{ca.}~4 to more than 12.\cite{Dshemuchadse2021} 
Similar interaction potentials had previously been found to assemble a one-component icosahedral quasicrystal, 
which is structurally completely distinct from known metallic quasicrystal structures: the observed structure exhibits local particle environments with 4--7 nearest neighbors, as opposed to metallic quasicrystals, which have high coordination numbers of 12 to 14, characteristic of sphere-packing-like geometries.\cite{Engel2015} 
In high-pressure simulations of a different series of isotropic pair potentials, other quasicrystals with mostly low coordination numbers were found. These quasicrystals exhibit octagonal, decagonal, and dodecagonal (\textit{i.e.}, 8-fold, 10-fold, and 12-fold) symmetries and also do not correspond to known crystal structures.\cite{Damasceno2017} 

In binary systems of octahedral colloidal clusters and spherical colloids, 
a new $cP6$-$A_2B_4$ structure type (space group $P\bar{4}3m$) was observed.\cite{Zanjani2016} 
In simulations of polydisperse particles, a complex Frank-Kasper phase $oS276$, which is also a decagonal quasicrystal approximant, was observed.\cite{Bommineni2019} 
In systems of hard polyhedra, more structures without atomic equivalents were found: a dodecagonal quasicrystal made up of tetrahedra,\cite{HajiAkbari2009} a ``$\beta$-tin''-like structure that constitutes a diamond-derivative structure $tI4$-$X$,\cite{Cersonsky2018} 
\textit{etc.}.\cite{Damasceno2012a} 

In studying the self-assembly behavior of particles that interact \textit{via} an isotropic pair potential with a $cF8$-C-forming (\textit{i.e.}, diamond-type) ground state,\cite{Marcotte2013} 
a variety of crystal structures were discovered that span both common densely packed structures, as well as crystals with low coordination numbers between 2 and 8.\cite{Bertolazzo2016} Some of these low-coordinated structures seem to correspond to those discovered by other computational studies ($hP2$-$X$, $tI4$-$X$, \textit{etc.}), while several are likely unique observations. 

It has been shown \textit{via} free-energy calculations that isotropic particles of two different sizes can stabilize a variety of structures if they are mixed, both with simple repulsive\cite{Travesset2015,Horst2016,Travesset2017,LaCour2019} 
as well as simple attractive interaction potentials.\cite{Ren2020} 
However, it should be noted that such ground-state calculations---similar to studies that investigate dense packings of different types of particles---do not necessarily reveal structures that are kinetically achievable \textit{via} assembly and will therefore not be discussed further here. 

As in simulations, a number of novel crystal structures have been discovered in experimental studies on the mesoscale in recent years. 
In some cases, the exact determination of their structure is challenging---by definition---due to the lack of points of comparison, paired with the sometimes limited structural information obtainable due to suboptimal crystallinity or limited sample sizes. 
Additionally, previously unknown structures are described in the literature in a variety of ways, which complicates matching them against reference data. 
In some cases, structures are reported directly \textit{via} their unit cells, as for example in the case of two new crystal structures found in a binary system of spherical and tetrahedral colloidal particles: a $cF36$-$AB_8$ phase (space group $Fm\bar{3}m$) and a $tP10$-$AB_4$ phase (space group $P\bar{4}m2$),\cite{Ducrot2017} 
both not equivalent to any known crystal structure types on the atomic scale. 
In a binary system of nanocrystals, an $hP34$-$A_9B_{25}$ 
structure (space group $P\bar{6}m2$) was observed.\cite{Boneschanscher2013} 
Also in binary nanoparticle systems, the structure type $cP14$-$AB_{13}$ ($Pm\bar{3}m$) was reported.\cite{Shevchenko2005,Shevchenko2006} 

Some structures are described by proxy through related structure types, as has been done in the case of $cI14$-$A_2B_{12}$ (space group $Im\bar{3}m$). 
This structure type was observed in binary systems of (DNA-functionalized) nanoparticles, as well as colloids,\cite{Ye2011,Macfarlane2011,Wang2015} and its structure was reported as being equivalent to $cI132$-Cs$_6$C$_{60}$, 
where one type of particle ($A$) takes on the positions of C$_{60}$ molecules in the unit cell. 
Although 
it has a relationship to a molecular structure type, the exact $cI14$ version had never been observed on the atomic scale---presumably due to the extreme size ratio between $A$- and $B$-type particles necessary to stabilize this arrangement. 

Another example of a structure description by proxy is the $cP54$-K$_4$Si$_{23}$ (or ``clathrate I'') derived structure found in DNA-functionalized nanoparticles shaped like trigonal bipyramids:\cite{Lin2017} 
here, each atomic position in the basic structure is actually occupied by a tetramer of trigonal bipyramids, with the nanoparticles effectively occupying the edges or bonds between the original atomic positions. Each atomic position in the Si-framework is connected to 4 other positions (with each bond being shared between 2 positions), leading to a particle count of $46 \times 4 / 2 = 92$ trigonal bipyramids per unit cell. The resulting $cP92$-$X$ structure type\cite{Lee2019} (space group $Fm\bar{3}n$) has no equivalent on the atomic scale.

Finally, deformable triblock Janus colloids self-assembled open structures, which were described \textit{via} their relationships to crystal structures with ionic bonding character:  
a scaffold of vertex-sharing tetrahedra, sometimes termed the ``pyrochlore'' structure, which consists of only the Cu-sites in the $cF24$-MgCu$_2$ Laves phase structure type or only the O-sites in the $cF24$-SiO$_2$ cristobalite structure type\cite{Li2021}---the same structure was also observed in simulations of particles interacting \textit{via} a repulsive square-shoulder pair potential.\cite{Pattabhiraman2017}
Triblock Janus colloids also formed a scaffold of vertex-sharing octahedra, which consists of only the O-sites of the $cP5$-CaTiO$_3$ perovskite structure type.\cite{Li2021} 


\section*{New properties on the mesoscale?}

Materials for a wide range of optical or photonic applications have been created from a variety of soft matter systems---block copolymers, colloids, nanoparticles, and biological materials.\cite{Gabinet2019} 
Self-assembled materials with photonic band gaps have been subject to in-depth investigations for many years, promising applications 
as elements of optical circuits, 
for improved light-harvesting, 
\textit{etc.} 
Due to the known wide and robust bandgap of the diamond structure, systems that could form this structure type have been studied intensively\cite{Ducrot2017,Wang2017,Cersonsky2018}
with only very recent experimental success.\cite{He2020} 
Complete photonic bandgaps were shown to exist in monoclinic crystal structures formed by dimer-shaped colloids,\cite{Hosein2010a,Hosein2010b,Fung2012} 
as well as several layered structures on square lattices and Archimedean $3^2.4.3.4$ tilings (to which the $tP30$-CrFe structure type can be mapped).\cite{Riley2012,Stelson2016,Stelson2017} 

Other ordered mesostructures were found to exhibit uncommonly high refractive indices and are predicted to enable cloaking capability.\cite{Kim2020} 
Crystalline mesostructures of spherical particles have potential for light-trapping applications.\cite{Marino2020} 
Mixed systems of colloids with distinct glass transition temperatures can be made into optical sensors based on time-dependent film formation properties.\cite{Schottle2021} 

The phononic bandstructures of mesostructures are governed by the shape of their building blocks, as well as their assembled crystal structures,\cite{Zanjani2015} 
and the combination of particles with different sizes can increase their phononic bandgaps significantly.\cite{Aryana2018} 

Colloidal semiconductor nanocrystals were found to exhibit strong interactions \textit{via} their organic ligands, 
leading to a pathway toward exploiting coherent phonons for light sources with high-frequency modulation.\cite{Poyser2016} 

Preferential alignment of nanoparticles along directions that have superior transport properties is necessary to create nanocrystal-derived thermoelectrics,\cite{Medlin2009} 
and it can be achieved by an assembly process that is directed \textit{via} external fields or by targeting anisotropic assembly structures.\cite{Kovalenko2015} 
If nanocrystals are assembled into epitaxially connected thin films, the resulting superior carrier transport properties can pave the way toward engineering new devices such as field-effect transistors.\cite{Zhao2021} 

By integrating colloidal nanocrystals with metallic, semiconducting, and insulating properties, high-performance devices such as field-effect transistors can be constructed that could be the basis for developing flexible, low-cost electronics.\cite{Choi2016,Kagan2019,Kagan2020,Liu2021} 
Potential device applications of nanocrystal mesostructures also include light-emitting diodes, photodetectors, solar cell components, and memory elements.\cite{Talapin2010} 

Future applications of nanocrystal assemblies will also explore their quantum optical properties, such as single-photon emission, optical and spin coherence, and spin-photon interfaces.\cite{Kagan2021} 
\textit{Via} nanotransfer printing, photonic and electronic devices can therefore be manufactured in a high-throughput manner, on flexible and curved surfaces, therefore creating opportunities for large-area metamaterial fabrication.\cite{Urbas2016,Paik2017} 
Patterning methods also allow for the fabrication of metamaterials that can be tuned in structure and function through a thermal trigger, eliciting a chiroptical response that will allow for the development of ultrathin lenses and polarizers.\cite{Guo2020} 


\section*{Search and design for new crystal structures}

A wide variety of approaches has been applied to explore the capacity of mesoscale building blocks to form different structures. 
While some studies target disordered structures\cite{Yu2021} 
or mesophases,\cite{Smalyukh2018,Kim2019} 
here we focus on the design of crystalline structures in particular. 

One important consideration in the context of targeting \textit{new} structures is that an otherwise powerful approach---that of inverse design---is \textit{not} the most promising path forward. 
Inverse design has been successful at targeting crystal structures that are \textit{known} from the atomic scale in a variety of mesoscale systems,\cite{Sherman2020} 
such as 
patchy particles,\cite{Long2018,Chen2018,Ma2019} 
faceted anisotropic shapes,\cite{vanAnders2015,Geng2019} 
and isotropic pair potentials.\cite{Torquato2009,Jain2014a,Adorf2018} 
However, since this pathway is based on previous knowledge of the targeted geometries, it explores only crystal structures that have been observed on the atomic scale and therefore are stabilized by highly discrete chemical elements and interactions. 
The vastly larger design space of structural building blocks on the mesoscale\cite{Glotzer2007,vanAnders2014b} enables us to strive for considerably more variety among structures and motifs, which means that our exploration of the structures that can be assembled in different kinds of systems can be equally broad. 

Over recent years, our ability to conduct increasingly extensive computational screenings for new structures has increased exponentially, with studies investigating for example the assembly behavior of hard polyhedra starting at discrete shapes\cite{HajiAkbari2009} and escalating to one- and two-dimensional parameter spaces in a matter of years.\cite{Damasceno2012b,Klotsa2018} 
Similarly, while earlier studies of the capability of isotropic pair potentials to form a variety of crystal structures had been focused on isolated findings,\cite{Zetterling2000,Elenius2009} recent studies have shown the acceleration of these investigations to screenings of increasingly large parameter spaces.\cite{Engel2015,Dshemuchadse2021} 
Current efforts are further expanding these approaches to higher-dimensional searches,\cite{Spellings2021} 
powered by machine-learning-enabled methods,\cite{Dai2020} 
and rendered possible by order parameters that permit the distinction of large varieties of crystal structures.\cite{Spellings2018,Reinhart2018} 

While it is challenging to investigate large swaths of parameter spaces or systems with many degrees of freedom \textit{via} computation, this is even more true for experimental studies, where each synthesized system and each set of parameters requires a far deeper commitment of laboratory and human resources. 
Several studies have investigated the structure formation of large varieties of particles (\textit{e.g.},\cite{Chanpuriya2016}), 
and some have begun to derive predictive rules for the stabilization of different structures: 
for binary systems of DNA-functionalized nanoparticles, for example, a set of design rules pertaining to their relative sizes was expressed 
that predicted which crystal structure would form from them
similar to Pauling's rules for ionic compounds.\cite{Macfarlane2011} 

The interplay of the thermodynamics of a particle system---and therefore its underlying energy landscape---and the kinetic contributions that enable the actual assembly of a structure 
is intricate and presumably varies drastically between different physical systems. 
The larger length scales of mesoscopic particle systems, as compared with microscopic ones, lead to an increase in assembly timescales, which in computational approaches are intrinsically limited. 
As a result, factors of particle kinetics play an increasingly large role in determining which structures are observed, and therefore need to be taken into account explicitly (by simulating full assembly trajectories \textit{vs.\ }comparing compound free energies) whenever possible. 
Polymorphism is a related obstacle, resulting from furrowed energy surfaces, \textit{i.e.}, the existence of multiple phases with competitive free energies, whose formation depends on assembly pathways, synthesis routes, and minute changes in crystal growth parameters. 
Here, too, the impact of kinetics has to be scrutinized in order to reliably document which circumstances lead to the assembly of a particular ordered structure.

\section*{Elucidate and describe new crystal structures}

The exploration 
and discovery of new crystal structures 
in soft condensed matter 
must 
be paired with reliable means of structure determination 
as well as the reporting of crystal structures through a common language. 
The elucidation of crystal structures on the mesoscale has been under constant development, enabling the determination of structures \textit{via} 
transmission electron microscopy with electron tomography\cite{Friedrich2009,Chen2010} 
and liquid cell transmission electron microscopy,\cite{Park2012} 
scanning probe microscopy and spectroscopy,\cite{Swart2016} 
as well as scattering experiments,\cite{Bian2011,Yager2014,Senesi2015} 
including small-angle X-ray scattering\cite{Li2016,Smilgies2021} 
(and electron diffraction, in particular for aperiodic structures\cite{Talapin2009}), 
and even \textit{in-situ} scattering\cite{Weidman2016} 
and electron microscopy.\cite{Liu2013,Sutter2016} 
In numerical simulations, structure determination is still mostly conducted on an \textit{ad-hoc} basis 
and the development of more robust and versatile tools, as well as the incorporation of standard crystallographic techniques still lies ahead. 

Another challenge to be tackled by the community is nomenclature: 
many of the known crystal structures are usually specified by a variety of names used for their hard-matter equivalents, ranging from prototypical compounds (``NaCl''), to the Strukturbericht designation (``A15 structure'') or other conventional descriptions originating from metallurgy (``$\sigma$ phase'') or applied mathematics (``cubic-close packing''). 
When new structures are described, their characteristics are often put into context by using related and known crystal structures as proxies. 
Sometimes space groups are reported, which is extremely useful but does not define a unique crystal structure comprehensively. Additionally, Wyckoff positions as well as all free parameters (of the unit cell and of the particle coordinates) need to be specified in order to render a full description of a crystal structure. 

Especially in referring to crystal structures \textit{via} short designations, a large variety of nomenclatures have been employed, ranging from the previously mentioned manners of referring to known, atomic structures to very specific abbreviations that are laid out separately and from scratch for each new study. 
In recent years, we have promoted the use of the above-mentioned Pearson symbol to serve as 
an information-rich shorthand to refer to structures both new and previously known: the notation captures both the symmetry and complexity of a crystal structure---\textit{via} the Bravais lattice and number of particles per unit cell---and simultaneously offers a more intuitive approach by linking these characteristics to a prototypical compound. 

The last component of the Pearson notation---a structural prototype---points to a larger purpose, challenge, and opportunity of reporting crystal structures in the field of soft condensed matter: 
by finding a common language that is shared between the disciplines of materials science, chemical engineering, physics, and chemistry, our collective explorations in this interdisciplinary field can be communicated in ways that enable us to identify overarching principles---matching identical or similar structures in disparate systems or conditions. 

One residual difficulty, however, is bound to persist despite the precise determination of particle positions and the adherence to a technically accurate structural description: 
the interpretation of crystal structures remains an ambiguous undertaking that relies on a correct abstraction of structural motifs and patterns. The meaning of structural models does not always transfer correctly between chemically distinct compounds---a condition that is inherently fulfilled in the length scale-spanning considerations in this manuscript. 
While the symmetry-based language to characterize fully ordered structures is precise, 
the geometrical description of crystal structures can remain ambiguous.
One can therefore inadvertently sidestep the more important underlying question---beyond ``where are the particles?'': 
``why are the particles arranged the way they are?''

\section*{Conclusions}

Over the past few years, a growing number 
of increasingly complex crystal structures has been discovered on the mesoscale. 
Experimentation and computation have both rendered a diverse set of structures in a variety of systems and with many types of interactions. 
On the path toward making functional colloidal materials, advances will now increasingly require the dedicated design and synthesis of more structures, including new geometries that have no precedent on any materials length scale. 

Simulation studies will need to stay ahead in their exploratory forays, to lighten 
the workload necessary for 
experimental studies through advancements in structure prediction, while also continually improving their relevance to physically meaningful interactions, 
and time scales. 
Experimentalists, on the other hand, are moving toward studies with higher throughput, while also expanding their methodological toolbox and the design dimensions along which the particles and their interactions can be tuned. 
By making headway in studies of both simulated and experimental systems, we will begin 
distilling a deeper understanding of when and how and why ordered crystals form from particles on the mesoscale. 
This will pave the way not just toward materials design of nano- and microstructured materials, 
but 
also a deeper understanding of emergent order and structure formation 
independent of length scale, 
broadening the impact of soft matter crystallography to become relevant for our understanding of atomistic crystallization and order.\cite{Kim2017}


\section*{Citation Diversity Statement}

Recent work in several fields of science has identified a bias in citation practices such that papers from women and other minority scholars are under-cited relative to the number of such papers in the field (\textit{e.g.},\cite{Caplar2017,Dworkin2020,Chatterjee2021}). In order to provide transparency and accountability for citation gender imbalance,\cite{Dworkin2020b} 
we used an open-source code that predicts the gender of the first and last author of each reference by using databases that store the probability of a first name being carried by a woman.\cite{Dworkin2020,Zhou2020} 
Excluding self-citations, 
our references contain 
8.18\% woman(first)/woman(last), 15.67\% man/woman, 10.09\% woman/man, and 66.06\% man/man 
by this measure. 
This method is limited in that \textit{(i)} names, pronouns, and social media profiles used to construct the databases may not, in every case, be indicative of gender identity and \textit{(ii)} it unfortunately cannot account for intersex, non-binary, or transgender people.

\section*{Acknowledgements}

The author thanks \textit{Journal of Applied Physics} Associate Editor Prof.\ LaShanda Korley for the nomination to submit this perspective article 
and Editor-in-Chief Dr.\ Andr\'e Anders for the invitation, 
as well as Dr.\ Erin G.\ Teich, Prof.\ Daphne Klotsa, and Dr.\ Thomas Dienel for helpful discussions.

\section*{Data Availability Statement}

Data sharing is not applicable to this article as no new data were created or analyzed in this study.

\section*{Author Declarations}

The author has no conflicts to disclose.

\appendix

\section*{Appendix}
\label{crystallographicnotationappendix}

The Pearson symbol is used throughout to denote crystal structure types. 
Its first component specifies a structure's Bravais lattice, of which there are 14: $cF$ = face-centered cubic, $cI$ = body-centered cubic, $cP$ = primitive cubic, $hP$ = hexagonal, $hR$ = rhombohedral, $tI$ = body-centered tetragonal, $tP$ = primitive tetragonal, $oF$ = face-centered orthorhombic, $oI$ = body-centered orthorhombic, $oS$ = base-centered orthorhombic, $oP$ = primitive orthorhombic, $mS$ = base-centered monoclinic, $mP$ = primitive monoclinic, $aP$ = primitive triclinic. 

The Bravais lattices are based on the 7 crystal systems: 
\begin{itemize}
\item $c$ = cubic ($a = b = c$, $\alpha = \beta = \gamma = 90^\circ$), 
\item $h$ = hexagonal ($a = b$, $\alpha = \beta = 90^\circ$, $\gamma = 120^\circ$), 
\item $t$ = tetragonal ($a = b$, $\alpha = \beta = \gamma = 90^\circ$), 
\item $o$ = orthorhombic ($\alpha = \beta = \gamma = 90^\circ$), 
\item $m$ = monoclinic ($\alpha = \gamma = 90^\circ$), 
\item $a$ = triclinic (no restrictions). 
\end{itemize}

The lattice centering types indicate how many lattice points are contained in one unit cell and where these are located:
\begin{itemize}
\item $P$ = primitive (1 lattice point): $(0,0,0)$,
\item $S$ = base-centered (2 lattice points): $(0,0,0)$ and either $(0,\frac{1}{2},\frac{1}{2})$ ($A$) or $(\frac{1}{2},0,\frac{1}{2})$ ($B$) or $(\frac{1}{2},\frac{1}{2},0)$ ($C$),
\item $I$ = body-centered (2 lattice points): $(0,0,0)$ and $(\frac{1}{2},\frac{1}{2},\frac{1}{2})$, 
\item $F$ = face-centered (4 lattice points): $(0,0,0)$ and $(0,\frac{1}{2},\frac{1}{2})$ and $(\frac{1}{2},0,\frac{1}{2})$ and $(\frac{1}{2},\frac{1}{2},0)$,
\item $R$ = rhombohedral (3 lattice points): $(0,0,0)$ and $(\frac{2}{3},\frac{1}{3},\frac{1}{3})$ and $(\frac{1}{3},\frac{2}{3},\frac{2}{3})$.
\end{itemize}

The number of atoms per unit cell given in the Pearson symbol refers to the indicated lattice centering and therefore does not necessarily correspond to the number of atoms per primitive unit cell (with the exception of the primitive Bravais lattices). 
All structures with rhombohedral space groups are given in their hexagonal setting. 

Given below are several examples for how this notation operates.
The diamond structure type $cF8$-C has space group $Fd\bar{3}m$ (no.\ 227) with 8 atoms per unit cell, positioned on the Wyckoff site $8a$ which (in origin choice 2) corresponds to positions $(\frac{1}{8},\frac{1}{8},\frac{1}{8})$ and $(\frac{7}{8},\frac{3}{8},\frac{3}{8})$, as well as---due to the face-centered lattice, whose vectors have to be added to these coordinates in all combinations---$(\frac{1}{8},\frac{5}{8},\frac{5}{8})$ and $(\frac{7}{8},\frac{7}{8},\frac{7}{8})$, $(\frac{5}{8},\frac{1}{8},\frac{5}{8})$ and $(\frac{3}{8},\frac{3}{8},\frac{7}{8})$, $(\frac{5}{8},\frac{5}{8},\frac{1}{8})$ and $(\frac{3}{8},\frac{7}{8},\frac{3}{8})$. 
All of these positions are occupied by C atoms in the prototypical diamond phase. 

The Frank-Kasper Z-phase structure type $hP7$-Zr$_4$Al$_3$ has space group $P6/mmm$ (no.\ 191) with 7 atoms per unit cell, positioned on Wyckoff sites 
$2c$, which corresponds to positions $(\frac{1}{3},\frac{2}{3},0)$ and $(\frac{2}{3},\frac{1}{3},0)$; 
$2e$, which corresponds to positions $(0,0,z)$ and $(0,0,\bar{z})$ (here with $z\approx0.3$); 
and $3f$, which corresponds to positions $(\frac{1}{2},0,0)$, $(0,\frac{1}{2},0)$, and $(\frac{1}{2},\frac{1}{2},0)$. 
In the prototypical compound Zr$_4$Al$_3$, sites $2c$ and $2e$ are occupied by Zr atoms, and site $3f$ is occupied by Al atoms. 
The aspect ratio of the unit cell of this structure type is $c/a \approx 1$. 

The L$1_0$ structure type $tP2$-CuAu has space group $P4/mmm$ (no.\ 123) with 2 atoms per unit cell, positioned on Wyckoff sites 
$1a$, which corresponds to position $(0,0,0)$, 
and $1b$, which corresponds to position $(0,0,\frac{1}{2})$. 
In the prototypical compound CuAu, site $1a$ is occupied by Au atoms and site $1b$ is occupied by Cu atoms.
The aspect ratio of the unit cell of this structure type is $c/a \approx 1.3$. 
(``L$1_0$'' is a ``Strukturbericht'' designation for intermetallic compounds.)

\section*{References}

\bibliographystyle{unsrt}
\bibliography{Refs}

\end{document}